\documentclass[twoside]{article}

\usepackage[accepted]{aistats2021}
\usepackage[round]{natbib}

\usepackage{amssymb,amsmath,amsthm,dsfont,xcolor}
\usepackage{mhequ}
\usepackage{mathabx}
\usepackage{accents,url}
\usepackage{hyperref}

\usepackage[]{graphicx}
\usepackage{caption}
\usepackage{subcaption}
\usepackage{multirow}

\usepackage{algorithm}
\usepackage[]{algpseudocode}
\def\algbackskip{\hskip-\ALG@thistlm}
\usepackage{setspace}
\usepackage{tikz}
\usetikzlibrary{shapes,snakes, bayesnet}

\def\m{\mathcal}
\def\mb{\mathbb}
\def\wt{\widehat}

\def\det{\mbox{det}}
\newcommand{\be}{\begin{equs}}
\newcommand{\ee}{\end{equs}}
\newcommand{\bpm}{\begin{pmatrix}}
\newcommand{\epm}{\end{pmatrix}}

\newcommand{\ind}{\mathbbm 1}
\newcommand{\bbE}{\mathbb{E}}

\DeclareMathOperator{\tr}{tr}

\def\mb{\mathbb}
\def\ind{\mathbbm{1}}

\newcommand{\abs}[1]{\left\vert#1\right\vert}

\def\ind{\mathds{1}}

\begin{document}



\twocolumn[

\aistatstitle{A Hybrid Approximation to the Marginal Likelihood}

\aistatsauthor{ Eric Chuu \And Debdeep Pati \And  Anirban Bhattacharya }

\aistatsaddress{ Texas A\&M University \And  Texas A\&M University \And Texas A\&M University } ]

\begin{abstract}
Computing the marginal likelihood or evidence is one of the core challenges in Bayesian analysis. While there are many established methods for estimating this quantity, they predominantly rely on using a large number of posterior samples obtained from a Markov Chain Monte Carlo (MCMC) algorithm. As the dimension of the parameter space increases, however, many of these methods become prohibitively slow and potentially inaccurate. In this paper, we propose a novel method in which we use the MCMC samples to learn a high probability partition of the parameter space and then form a deterministic approximation over each of these partition sets. This two-step procedure, which constitutes both a probabilistic and a deterministic component, is termed a Hybrid approximation to the marginal likelihood. We demonstrate its versatility in a plethora of examples with varying dimension and sample size, and we also highlight the Hybrid approximation's effectiveness in situations where there is either a limited number or only approximate MCMC samples available. 
 
\end{abstract}

\section{INTRODUCTION} \label{sec:intro}

%%%%%% intro.tex
Model selection and model averaging are among the most important inferential goals in Bayesian statistics. These goals inherently rely on evaluating model uncertainty, which in turn comes down to calculating the marginal likelihood of competing models. This makes accurate and efficient computation of the marginal likelihood an important problem.

Suppose we observe data $y$ with a likelihood function $p \left(y \mid u \right)$ indexed by $u$ coming from some parameter space $\mathcal{U}$. Provided that the prior distribution over the unknowns is specified, the marginal likelihood or evidence can be written as
\begin{equation}  \label{eq:ml_general}
	p\left(y\right) =  \int_{\mathcal{U}} p\left(y \mid u\right) p\left(u\right) \, du.
\end{equation}
Barring specific conjugate settings, the marginal likelihood is analytically intractable in practice and poses a computationally challenging problem. Since numerical integration becomes infeasible beyond moderate dimension, Monte Carlo approximations present an alternative solution. In much of the literature devoted to estimating this quantity, the recurring idea is to form an asymptotically unbiased approximation of (\ref{eq:ml_general}) using MCMC samples. However, running an MCMC algorithm in addition to forming a Monte Carlo approximation can quickly accrue error as the dimension of the parameter space increases. Consequently, these algorithms may require an exceedingly large number of samples in order to form accurate estimates. In many problems, however, obtaining MCMC samples is time-consuming and potentially unreliable. As such, the need for an approximation that does not too heavily rely on both the quantity and quality of the MCMC samples is evident.

Some commonly used algorithms include Laplace's method \citep{tierney_kadane_1986}, which assumes that the posterior distribution can be approximated with a normal distribution, and the Harmonic Mean estimator \citep{newton_raftery_1994}, which is easy to implement, but has been shown to be unstable and can have infinite variance \citep{newton_raftery_1994}, \citep{raftery_newton_satagopan_krivitsky_2007}. On a similar vein, the Adjusted Harmonic Mean estimator \citep{lenk_2009} and the Corrected Arithmetic Mean estimator \citep{pajor_2017} leverage the harmonic mean and arithmetic mean identities, respectively, with the idea of sampling from high posterior probability regions of the parameter space to improve upon the original estimators. Annealed Importance Sampling \citep{neal_2001} uses a dynamic importance sampling function that sequentially transitions through intermediate distributions to the target distribution.

Other popular algorithms include Chib's method \citep{chib_1995,chib_jeliazkov_2001}, Bridge Sampling \citep{meng_wong}, Warp Bridge Sampling \citep{meng_schilling_2002}, and Nested Sampling \citep{skilling_2006}. See \cite{friel_wyse_2012} for a more comprehensive overview and discussion of these algorithms. There have also been recent developments in variational inference techniques that provide alternative ways to approximate and bound the marginal likelihood \citep{pmlr-v37-rezende15, pmlr-v37-salimans15}. 

In contrast to these methods, we propose a novel approach which can be thought of as a hybrid between probabilistic and deterministic procedures. A high level view of our method can be broken down into two major steps: (i) the MCMC samples are used to learn a partition of the parameter space $\m U$, and (ii) with this partition, we then make a deterministic approximation to the log posterior on each of the partition sets. In essence, we seek to exploit the assumption that the posterior distribution will be far from a uniform looking distribution and instead show concentration around some parameter. If the partition obtained from the MCMC samples can identify areas of high posterior mass by carving up these regions more finely, then we are better equipped to make an accurate approximation to the log posterior over each of these regions. Given the use of a probabilistic procedure in step (i), coupled with a deterministic calculation in step (ii), we refer to the resulting approximation to the marginal likelihood as the \textit{Hybrid estimator}. 

Our contribution fundamentally provides a way to bypass the need for a large number of posterior samples for accurate computation of the marginal likelihood. In many applications, evaluating the likelihood can be extremely time consuming, so in turn, collecting lots of posterior samples in such cases is prohibitively expensive in both time and computation. The typical guarantees for MCMC-based estimates of the marginal likelihood are asymptotic in the number of posterior samples. Our approach instead only uses the MCMC samples to learn a skeleton of the posterior distribution, which then simplifies the subsequent calculation; hence, the Hybrid estimator establishes a scalable framework for computing the evidence in high dimensional problems.

The paper is organized as follows. In Section \ref{sec:methods}, we motivate each step in the algorithm and provide a formal statement of the Hybrid approximation scheme. In Section \ref{sec:results}, we demonstrate the performance of the Hybrid estimator in a variety of simulation studies and compare the results with some of the aforementioned estimators. Finally, in Section \ref{sec:conclusion}, we briefly discuss some details for extensions and future work.

\section{METHODOLOGY} \label{sec:methods}

We introduce some preliminary notation. Let $\gamma$ be a probability density with respect to the Lebesgue measure on $\mb R^d$ given by 
\begin{align*}
\gamma(u) &= \frac{e^{-\Phi(u)} \, \pi(u)}{\m Z}, \quad u \in {\m U} \subseteq \mb R^d. 
\end{align*}
When $\Phi(\cdot)$ corresponds to a negative log-likelihood function and $\pi(\cdot)$ a prior distribution, $\gamma(\cdot)$ is the corresponding posterior distribution, although such an interpretation is not necessary for our approach. Then, the marginal likelihood has the following form,
\begin{align} \label{eq:Z}
    \m Z = \int_{\m U} e^{-\Psi(u)} \, du,
\end{align}
where $\Psi(u) = \Phi(u) + (-\log \pi(u))$ is the negative log-posterior. As stated before, while we can evaluate $\Psi$, we are unable to compute the integral in (\ref{eq:Z}). We can address this problem using the two sub-routines mentioned in the previous section. First, we find a partition of the parameter space that gives more attention to (i.e, more finely partitions) regions of the posterior that have high posterior mass. Next, we propose a suitable approximation for $\Psi$ that allows for easier evaluation of the integral over each of the partition sets learned from the previous step. These steps used in conjunction with each other give us a way to approximate $\m Z$ by computing a simplified version of the integral over partition sets of the parameter space that have ideally taken into account the assumed non-uniform nature of the posterior distribution.

\subsection{Deterministic Approximation} \label{sub:ii}
We first elaborate on our strategy to replace $\Psi$ with an approximation $\widehat{\Psi}$. Our starting point is the following observation: fix $q \in (0, 1)$ small and let $A \subseteq \m U$ be a compact subset with $\gamma(A) \ge (1 - q)$. Rearranging this equation, one obtains $(1 - q) \le \gamma(A) = \m Z^{-1} \int_A e^{-\Psi(u)} du \le 1$, leading to the two-sided bound
\begin{align}\label{eq:two_sided}
\int_{A} e^{-\Psi(u)} \, du \le \m Z \le \frac{1}{1-q} \, \int_{A} e^{-\Psi(u)} \, du. 
\end{align}
We then make the following approximation
\begin{align}\label{eq:init_approx}
\log \m Z \approx F_A :\,= \log \bigg[\int_A e^{-\Psi(u)} \, du \bigg]. 
\end{align}
From Eq.\,\eqref{eq:two_sided}, it is immediate that $|\log \m Z - F_A| \le \log\{1/(1-q)\} \approx q$ for $q$ small. Henceforth, we aim to estimate the quantity $F_A$. This initial approximation step can be thought of as compactifying the parameter space to reduce its entropy. Even if $\m U$ itself is compact, $\gamma$ can be highly concentrated in a region $A$ with $\mbox{vol}(A) \ll \mbox{vol}(\m U)$, particularly when the posterior exhibits concentration \citep{ghosal2007convergence}, and it is judicious to eliminate such low posterior-probability regions.

Having compactified the integral domain, our general plan is to replace $\Psi$ with a suitable approximation $\widehat{\Psi}$ on the compact set $A$. In this article, we specifically focus on a piecewise constant approximation of the form 
\begin{align} \label{eqn:constant_approx}
\wt{\Psi}(u) = \sum_{k=1}^K c_k^\star \cdot \ind_{A_k}(u), 
\end{align}
where $\m A = \{ A_1, \ldots, A_K \}$ is a partition of $A$, i.e., $A = \bigcup_{k=1}^K A_k$ and $A_k \cap A_{k'} = \emptyset$ for all $k \ne k'$, and $c_k^\star$ is a representative value of $\Psi$ within the partition set $A_k$. To simplify the ensuing calculations, we further restrict ourselves to dyadic partitions in this article so that each of the partition sets is rectangular, $A_k = \prod_{l=1}^d [a_k^{(l)}, b_k^{(l)}]$. This leads to the approximation 
\begin{align} \label{eqn:ml_approx} 
\int_A e^{-\Psi(u)} \, du \approx \int_A e^{-\wt{\Psi}(u)} \, du = \sum_{k=1}^K e^{-c_k^\star} \cdot \mu(A_k),
\end{align}
where $\mu(B) = \int_B 1 \, du$ denotes the $d$-dimensional volume of a set $B$. We eventually define 
{\small\begin{align}\label{eq:Fhat}
\wt{F}_A :\,= \log \Big[\int_A e^{-\wt{\Psi}(u)} \, du \Big] = \log \bigg[\sum_{k=1}^K e^{-c_k^\star} \cdot \mu(A_k) \bigg]
\end{align}}\\[1ex]
to be our estimator of $F_A$, and hence of $\log \m Z$. The choice of the piecewise constant approximation is motivated both by its approximation capabilities \citep{binev_cohen_dahmen_devore_temlyakov_2005} as well as the analytic tractability of the approximating integral in Eq.\,\eqref{eq:Fhat}. We remark here that the integral remains tractable if a piecewise linear approximation is employed, suggesting a natural generalization of our estimator.  

Since $F_A$ is a non-linear functional of $\Psi$, it is reasonable to question the validity of the approximation in Eq.\,\eqref{eqn:ml_approx}, or equivalently, the approximation of $F_A$ with $\wt{F}_A$ --- even if $\wt{\Psi}$ is a good approximation to $\Psi$, it is not immediately clear if the same should be true of $\wt{F}_A$. Using an interpolation trick, we show below that the approximation error $|\wt{F}_A - F_A|$ can be bounded in terms of a specific distance between $\wt{\Psi}$ and $\Psi$. Define 
\begin{align*} 
F(t) = \log \bigg[\int_A e^{ -\big(t \Psi(u) + (1-t) \widehat{\Psi}(u) \big) } \, du \bigg], \quad t \in [0, 1]. 
\end{align*}
Clearly, $F(0) = \widehat{F}_A$ and $F(1) = F_A$, so that 
\begin{align*}
F_A - \widehat{F}_A = F(1) - F(0) = \int_0^1 F'(t) dt. 
\end{align*}
Computing $F'$, we get
\begin{align*}
F'(t) 
& = \frac{ - \int_A \big(\Psi(u) - \widehat{\Psi}(u) \big)\, e^{ -\big(t \Psi(u) + (1-t) \widehat{\Psi}(u) \big) } \, du }{ \int_A e^{ -\big(t \Psi(u) + (1-t) \widehat{\Psi}(u) \big) } \, du} \\
& = - \bbE_{U \sim \pi_t} \big(\Psi(U) - \widehat{\Psi}(U)\big),
\end{align*}
where $\pi_t$ is the probability density on $A$ given by 
\begin{align*}
\pi_t(u) \,\propto\, e^{ -\big(t \Psi(u) + (1-t) \widehat{\Psi}(u) \big) }, \quad u \in A. 
\end{align*}
Using the integral representation, we can now bound the approximation error,
\begin{align*} 
|F_A - \widehat{F}_A| \le \sup_{t \in [0, 1]} \big| \bbE_{U \sim \pi_t} \big(\Psi(U) - \widehat{\Psi}(U)\big) \big|. 
\end{align*}
Interestingly, note that $\pi_1 \,\propto\, \gamma \ind_A$ is our target density restricted to $A$, and $\pi_0(u) \,\propto \, e^{-\wt{\Psi}(u)} \, \ind_A(u)$ has normalizing constant $\wt{F}_A$. The collection of densities $\{\pi_t\}$ can therefore be thought of as continuously interpolating between $\pi_0$ and $\pi_1$. Piecing together the various approximations, we arrive at the following result. 

{\bf Proposition 1.} {\em For any compact subset $A \subseteq \m U$, we have}
{\small\begin{align*}
|\wt{F}_A - \log \m Z| \le \sup_{t \in [0, 1]} \big| \bbE_{U \sim \pi_t} \big(\Psi(U) - \widehat{\Psi}(U)\big) \big| + \log \bigg( \frac{1}{\nu(A)} \bigg).    
\end{align*}}

Here, $\nu$ denotes the Lebesgue measure on $\mathbb{R}^D$. The first term in the right hand side above can be further bounded by $\|\Psi - \wt{\Psi}\|_{\infty} :\,= \sup_{u \in A} |\Psi(u) - \wt{\Psi}(u)|$. This conclusion is not restricted to the piecewise constant approximation and can be used for other approximations, such as the piecewise linear one.

\subsection{High Probability Partitioning of the Parameter Space} \label{sub:i}

\begin{figure*}[!ht]
    \centering
    \newlength{\MyHeight}
    \settoheight{\MyHeight}{\includegraphics[scale=0.4]{example-image-a}}
    \begin{subfigure}[t]{0.49\textwidth}
        \centering
        \includegraphics[scale=0.26]{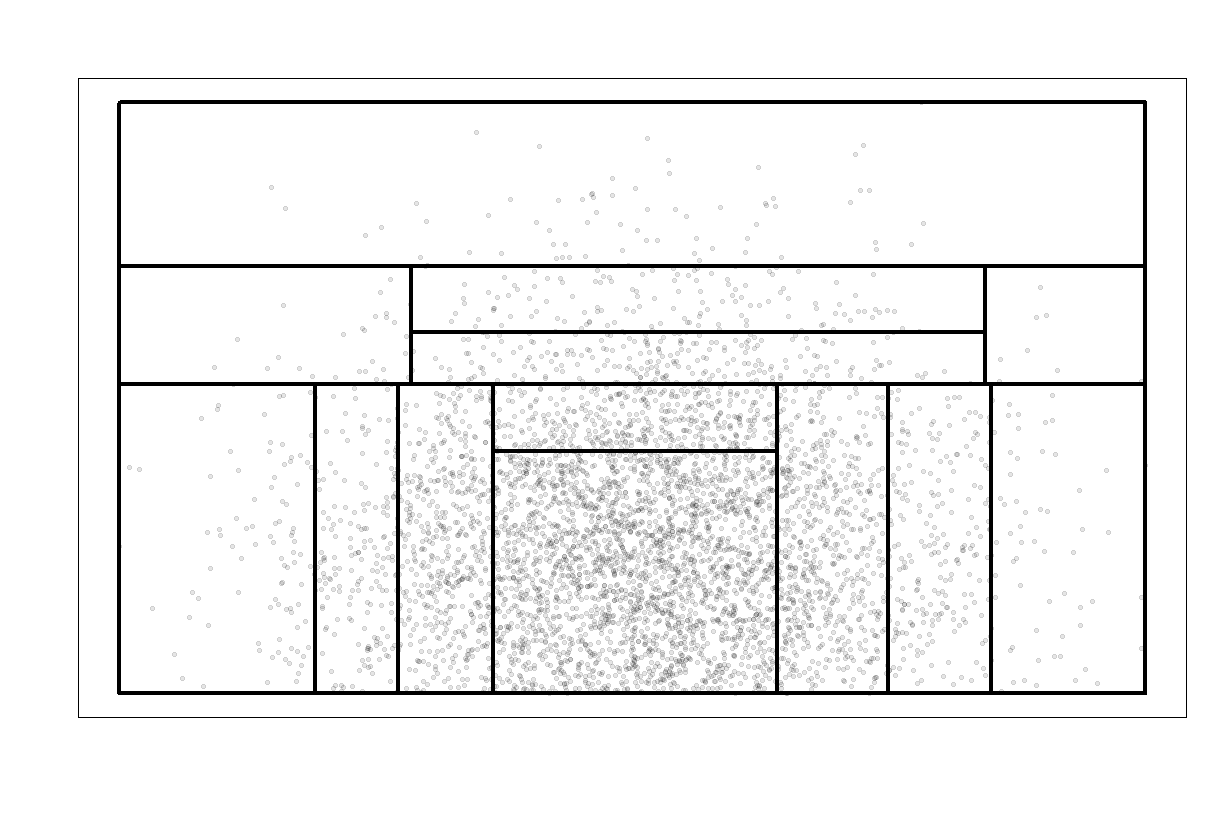}
    \end{subfigure}
    \hfill
    \begin{subfigure}[t]{0.49\textwidth}
        \centering
        \includegraphics[scale=0.26]{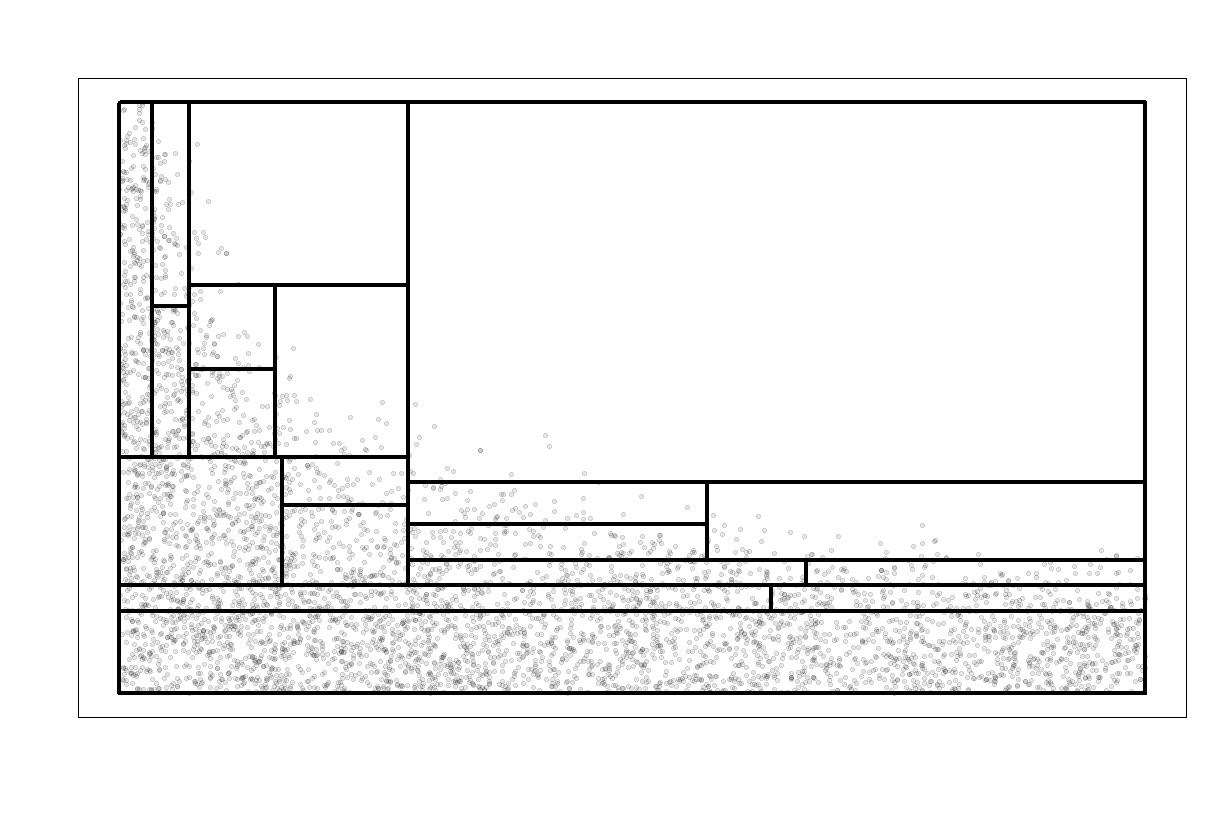}
    \end{subfigure}
    \caption{\textit{Left: bivariate normal distribution truncated to the first orthant. Right: A density of the form $\gamma(u) \ \propto \ \exp(-nu_1^2u_2^4) \pi(u)$, where $u \in [0,1]^2$ and $\pi(\cdot)$ is the uniform measure on $[0,1]^2$. For this simulation, $n = 1000$. Both plots have 5000 MCMC samples with the partition returned from fitting a CART model.}\label{fig:twod_partition}}
\end{figure*}

Next, we address the task of obtaining a suitable partition of the parameter space. Clearly, traditional quadrature methods would render this method ineffective, requiring the number of function evaluations to grow exponentially with $d$. 
Furthermore, with a posterior distribution that exhibits any degree of concentration, there will indubitably be regions of $\m U$ where the posterior probability is close to 0. From a computationally mindful standpoint, it makes sense to then focus on more finely partitioned regions of $\m U$ that have high posterior probability. With this in mind, we turn to using samples from $\gamma$ to obtain such a partition. Specifically, let $u_1, \ldots, u_J$ be approximate samples from $\gamma$, e.g., the output of an MCMC procedure. 
We treat $\{ (u_j, \Psi(u_j) \}_{j=1}^J$ as covariate-response pairs and feed them to a tree-based model such as CART \citep{breiman_1984}, implemented in the \texttt{R} package \texttt{rpart} \citep{rpart_package}, to obtain a dyadic partition. While the MCMC samples are typically used to construct Monte Carlo averages, we use them to construct a high probability partition of the parameter space. We assume the capability to evaluate $\Psi$, which is a very mild assumption since obtaining samples from $\gamma$ using even a basic sampler like Metropolis--Hastings requires evaluating $\Psi$. Finally, the above procedure implicitly suggests the compactification $A$ to be a bounding box using the range of posterior samples, $A = \otimes_l \left[ \min \{ u^{(l)}_{j} \}, \max \{ u^{(l)}_{j} \} \right]$, $\ 1 \leq j \leq J$, $1 \leq l \leq d$, where $u^{(l)}_{j}$ is the $l$th component of $u_{j}$.

\subsection{Partitioning in Two Dimensions} \label{sec:two_dim}

Before moving into higher dimensions, we provide an illustration of the process described in the previous section in 2 dimensions, where the partitioning can be easily visualized. Suppose $\gamma$ is a density on $\mathbb{R}^2$ supported on $\m U \subseteq \mathbb{R}^2$, and $u_j \sim \gamma$ for $j = 1, \ldots, J$. Forming the pairs, $\{(u_j, \Psi(u_j))\}_{j=1}^J$, we then fit a CART model to these points and extract the decision rules, which form a dyadic partition of the aforementioned bounding box $A \subseteq \m U$. Denote the partition as $\m A = \{ A_1, \ldots, A_K \}$. Plotting the sampled points and overlaying the partitions learned from the regression tree, we observe in Figure \ref{fig:twod_partition} that areas of $\m U$ with a high concentration of points coincide with regions that are more finely partitioned by the regression tree. Taking $\gamma$ to be a posterior distribution, we see that this behavior of partitioning areas of greater posterior mass is desirable in producing a better approximation. Equipped with the partition $\m A$, we need only to determine the representative point of each partition set in order to form the approximation to $\Psi$.

Recall that CART fits a constant for each point within a given partition set. At any given stage, the CART model will search for the optimal predictor value, $u = (u_1, u_2)$, on which to partition the remaining points such that the sum of squares error (SSE) between the response, $\Psi(u)$, and the predicted constant is minimized. In particular, to partition data into two regions $A_1$ and $A_2$, the objective function is given as 
\begin{align} \label{eq:SSE}
    & SSE = \sum_{u_i \in A_1} (\Psi(u_i) - c_1)^2 + \sum_{u_i \in A_2} (\Psi(u_i) - c_2)^2.
\end{align}
Upon minimization of the SSE, the resulting partition sets $A_1$ and $A_2$ have fitted values $c_1$ and $c_2$, respectively. For each partition set $A_k \in \m A$, a natural choice for the representative point $c_k^\star$ is the fitted value for $A_k$ produced by the tree-fitting algorithm. Following this two-step process of using CART to obtain both the partition and the fitted values for each of the partition sets and then plugging these into Eq. (\ref{eqn:ml_approx}), we obtain the Hybrid approximation to the marginal likelihood.

\subsubsection{Conjugate Normal Model} \label{sec:conjugate2d}

We consider the following conjugate normal model: $y_{1:n} \mid \mu, \sigma^2 \sim \m N(\mu, \sigma^2)$, $\mu \mid \sigma^2 \sim \m N(m_0, \sigma^2 / w_0)$, $\sigma^2 \sim \m{IG}(r_0 / 2, s_0 / 2)$, where $\m{IG}(\cdot, \cdot)$ denotes the inverse-gamma distribution. In order to compute the Hybrid estimator, we require samples from the posterior distribution and a way to evaluate $\Psi$. In this example, the posterior distribution of $u = (\mu,\sigma^2)$ is known, and since the likelihood and prior are specified, the evaluation of $\Psi$ is straightforward. With this architecture in place, we feed the pairs, $\{ (u_{j}), \Psi(u_{j}) \}_{j=1}^J$, through CART to obtain a partition over the parameter space and each partition set's representative point. Then, we use Eq.\,(\ref{eqn:ml_approx}) to compute the final approximation.
\begin{table}[h]
\caption{\textit{Normal Inverse-Gamma Example. We report the mean, standard deviation, average error (AE, truth - estimated), and the root mean squared error (RMSE), taken over 100 replications. Each replication has 50 observations and 1000 posterior samples. The true log marginal likelihood is -113.143. Estimators include the Harmonic Mean estimator (HME), Corrected Arithmetic Mean estimator (CAME), Bridge Sampling estimator (BSE), and the Hybrid estimator (HybE).}}
\label{tab:nig2d}
\begin{center}
\resizebox{\columnwidth}{!}{%
\begin{tabular}{l|c|c|c|c}
\textbf{Method}  & \multicolumn{1}{l|}{\textbf{Mean}} & \multicolumn{1}{l|}{\textbf{SD}} & \multicolumn{1}{l|}{\textbf{AE}} & \multicolumn{1}{l}{\textbf{RMSE}} \\ \hline
log HME          & -104.762                    & 0.733                    & -8.381                    & 8.431      \\
log CAME         & -112.704                    & 0.048                    & -0.439                    & 0.441      \\
log BSE          & -113.143                    & 0.006                    & 0                         & 0.006      \\
log HybE         & -113.029                    & 0.025                    & -0.114                    & 0.117                            
\end{tabular}%
}
\end{center}
\end{table}

Table \ref{tab:nig2d} shows results for the Hybrid estimator and a number of other competing methods. Here, the true log marginal likelihood can be computed in closed form, so we have direct comparisons to the ground truth. All estimators except for the Harmonic Mean estimator give accurate approximations to the log marginal likelihood.

\subsection{Algorithm Description}
Until this point, the representative point within each partition has simply been the fitted value for each partition returned from the CART model. When $\{(u_j , \Psi(u_j))\}_{j=1}^J$ is fed into the tree, it attempts to optimize the sum of squared errors as in Eq.\,\eqref{eq:SSE}. Note, however, that our eventual objective is to best approximate the functional $\log \int_A e^{-\Psi}$, and it is not unreasonable to suspect that the optimal value for $A_k$ chosen by the regression tree model may not be a suitable choice for our end goal, especially for higher dimensions. Simulations in higher dimensions indeed confirm this. Before suggesting a remedy, we offer some additional understanding into the approximation mechanism that guides us toward an improved choice. To that end, write $\wt{F}_A$ from Eq.\,\eqref{eq:Fhat} as  
\begin{align*}
    \widehat{F}_A = \log \bigg[\sum_{k=1}^K e^{-c_k^\star} \, p_k \bigg] + \log \mu(A) 
    :\, = \widehat{G} + \log \mu(A),
\end{align*}
where recall $\mu(B) = \mbox{vol}(B)$ is the Lebesgue measure of a Borel set $B$, and we define $p_k :\,= \mu(A_k)/\mu(A)$. We can also write $F_A = G + \log \mu(A)$, with 
\begin{align*}
    G &:\,= \log \bigg[ \frac{1}{\mu(A)} \int_A e^{- \Psi(u) } du \bigg] \\
    &= \log \bigg[\sum_{k=1}^K p_k \frac{1}{\mu(A_k)} \int_{A_k} e^{- \Psi(u) } du \bigg] \\
    &= \log \bigg[ \sum_{k=1}^K e^{-c_k} \, p_k \bigg],
\end{align*}
where 
\begin{align*}
	e^{-c_k} = \frac{1}{\mu(A_k)} \int_{A_k} e^{- \Psi(u) } du = \bbE_{U_k \sim \mbox{Unif}(A_k)} \big[e^{-\Psi(U_k)} \big].
\end{align*}
Thus, for $\wt{G}$ to approximate $G$, we would ideally like to have each $c_k^\star$ chosen so that $e^{-c_k^\star}$ targets $e^{-c_k}$. Importantly, the above exercise suggests the appropriate scale to perform the approximation -- rather than working in the linear scale as in Eq.\,\eqref{eq:SSE}, it is potentially advantageous to work in the exponential scale.

\subsubsection{Choosing the Representative Point}

Based on the above discussion, we define a family of objective functions
\begin{align}\label{eq:obj_Q}
    Q_k(c) = \sum_{u \in A_k} \frac{|e^{-\Psi(u)} - e^{-c}|}{e^{-\Psi(u)}}, \quad c \in A_k,
\end{align}
one for each partition set $A_k$ returned by the tree, and set $c_k^\star = \mathrm{argmin}_{c} Q_k(c)$. We experimented with a number of different metrics before zeroing in on the above relative error criterion in the exponential scale. Minimizing \eqref{eq:obj_Q} is a weighted $\ell_1$ problem and admits a closed-form solution. 

Thus, our overall algorithm can be summarized as follows. We obtain samples $u_1, \ldots, u_J$ from $\gamma$, and feed $\{(u_j, \Psi(u_j)\}_{j=1}^J$ through a tree to partition the bounding box $A$ of the samples. Then, rather than using the default fitted values returned by the tree, we take the representative value $c_k^\star$ within each $A_k$ as the minimizer of $Q_k$. These $c_k^\star$s are then used to compute $\wt{F}_A$ as in \eqref{eq:Fhat} -- note that $\wt{F}_A$ can be stably computed using the log-sum-exp trick. Finally, we declare $\wt{F}_A$ as $\log \wt{\m Z}$, our estimator of $\log \m Z$.

    \begin{algorithm}
    \caption{Hybrid Approximation}\label{alg:hybrid}
    \hspace*{\algorithmicindent} \textbf{Input:} Sampler for $\gamma$, method for evaluating $\Psi$ \\
    \hspace*{\algorithmicindent} \textbf{Output:} Estimate of the log marginal likelihood
    \begin{algorithmic}[1]
    \State Sample  $\{u_j\}_{j=1}^J$ from $\gamma$
    \State Fit $\{ (u_j, \Psi(u_j)) \}_{j=1}^J$ using a regression tree
    \State From the fitted tree, extract the dyadic partition $\mathcal{A} = \{ A_1, A_2, \ldots, A_K\}$ of the bounding box $A = \otimes_l \left[ \min \{ u^{(l)}_{j} \}, \max \{ u^{(l)}_{j} \} \right]$ determined by the samples, with $A_k = \prod_{l=1}^d [a_k^{(l)}, b_k^{(l)}]$
    \For{$k \in \{ 1 , \ldots,  K \}$}
        \State $c_k^\star \gets \mathrm{argmin}_{c \in A_k} \log Q_k(c)$
        \State $\wt{\mathcal{Z}}_k \gets e^{-c_k^\star} \ \prod_{l=1}^d \big(b_k^{(l)} - a_k^{(l)}\big)$
    \EndFor
    \State Use the log-sum-exp trick to compute the final estimator, $\log \wt{\mathcal{Z}} = \texttt{log-sum-exp} \left( \log \hat{\mathcal{Z}}_1, \ldots, \log \hat{\mathcal{Z}}_K  \right)$
    \end{algorithmic}
    \end{algorithm}

\section{RESULTS} \label{sec:results}

%%%% truncated normal linear regression

\begin{figure*}[!ht]
    \centering
    \settoheight{\MyHeight}{\includegraphics[scale=0.49]{example-image-a}}
    \begin{subfigure}[t]{0.49\textwidth}
        \centering
        \includegraphics[scale=0.32]{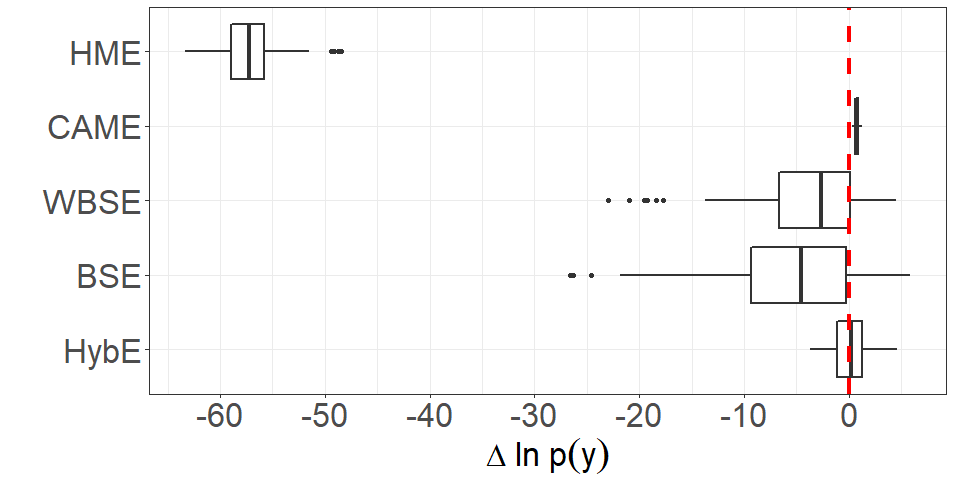}
    \end{subfigure}
    \hfill
    \begin{subfigure}[t]{0.49\textwidth}
        \centering
        \includegraphics[scale=0.32]{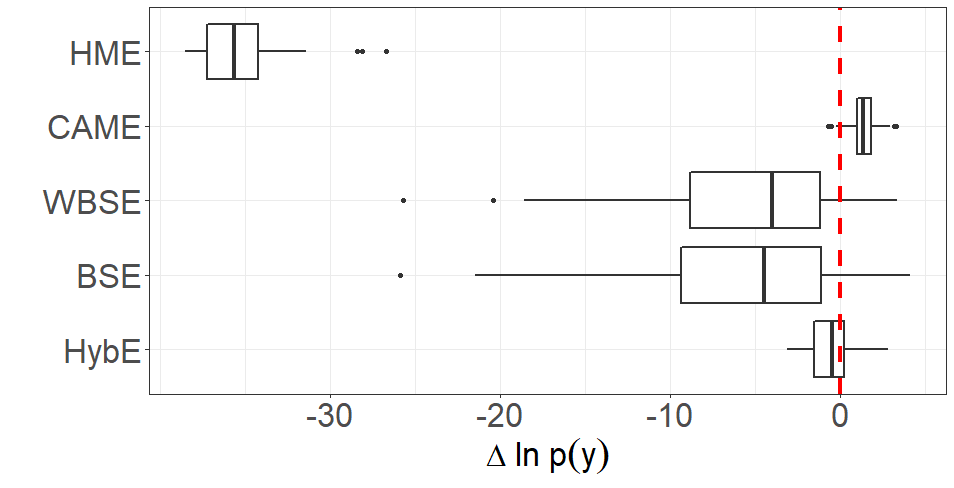}
    \end{subfigure}
    \caption{\textit{Boxplots of the error (truth - estimate) for the log marginal likelihood in the MVN-IG (left, true $\log p(y)$: -303.8482) and truncated MVN (right, true $\log p(y)$: -250.2755) examples. Both examples correspond to 20-dimensional parameter spaces. Results are reported over 100 simulations, with 100 observations. Estimates are based on 45 MCMC samples. The results correspond to the natural logarithm of each of the estimators. \label{fig:regression}}}
\end{figure*}
 
 % The true log marginal likelihood in both is: -250.2755 (calculated using $\texttt{TruncatedNormal}$ package.)

In the following experiments, we present a variety of problem settings. First, we consider the linear regression model under different prior specifications where the true marginal likelihood is known, so we can easily verify the accuracy of any subsequent approximations. We then extend the application of the Hybrid estimator to examples for which the parameter is a $d \times d$ covariance matrix, thus showcasing its versatility even when the parameter space is non-Euclidean. We examine the performance of the Hybrid estimator alongside competing methods and focus primarily on situations where the posterior samples are either few in number or non-exact. In addition to the Hybrid estimator (HybE), we examine the following additional estimators: Bridge Sampling estimator (BSE), Warp Bridge Sampling estimator (WBSE), Harmonic Mean estimator (HME), and Corrected Arithmetic Mean estimator (CAME). The BSE and WBSE results are obtained using the \texttt{bridgesampling} package \citep{bs_package}. Corresponding calculations and formulae for posterior parameters and analytical marginal likelihoods are given in the Supplement.

\textcolor{black}{We emphasize that in the experiments provided in this section we seek to mimic scenarios where posterior sampling is highly expensive and/or mixing is poor. By considering a small number of samples as the input for these marginal likelihood estimation algorithms, we provide a realistic scenario for the regime in which we wish to operate. In the examples in Sections \ref{sec:regression} and \ref{sec:graphical}, we sample directly from exact posterior distribution, while in Section \ref{sec:vb}, we use approximate samples from the posterior distribution.}

\subsection{Bayesian Linear Regression} \label{sec:regression}
Consider the following setup of the linear regression model, $y = X\beta + \varepsilon$, where $y \in \mathbb{R}^n$, $X \in \mathbb{R}^{n \times d}$, $\beta \in \mathbb{R}^d$, and $\varepsilon \sim \m N(0, \sigma^2 I_n)$. In the next two examples, we consider different prior distributions on $\beta$ and $\sigma^2$.

\subsubsection{Multivariate Normal Inverse-Gamma Model} \label{sec:mvnig-reg}    % page 77 goodnotes

We assume a multivariate normal inverse-gamma (MVN-IG) prior on $(\beta, \sigma^2)$, where $\beta \mid \sigma^2 \sim \m N_d ( \mu_{\beta}, \sigma^2 V_{\beta})$, $\sigma^2 \sim \m{IG} (a_0, b_0)$. Given this choice of the prior, the posterior distribution is known to be $\beta \mid \sigma^2, y \sim \m N \left( \mu_n, \sigma^2 V_{n} \right)$, $\sigma^2 \mid y \sim \m{IG} \left( a_n, b_n \right)$. In our simulation, we take $d = 19$, so that $u = (\beta, \sigma^2) \in \mathbb{R}^{20}$. Since the log marginal likelihood in this example is well known, we evaluate each of the estimates against the true value. In Figure \ref{fig:regression}, we plot the errors for each of the estimators when only 45 MCMC samples are used for each approximation. The accuracy and standard error of the Hybrid estimator are clearly superior compared to the well-established estimators. 

%We note here that previous accuracy issues that arose when increasing the dimension of the parameter space have been resolved.

\iffalse
\begin{figure}[!h]
\includegraphics[scale=0.32]{figs/mvnig_d20_j5e3.png}
\caption{\textit{Boxplots of the error (truth - estimate) for the log marginal likelihood MVN-IG examples. Here $u = (\beta, \sigma^2) $ is a 20-dimensional parameter. Results are reported over 100 simulations, with 100 observations and 5000 samples from the posterior.} \label{fig:reg-mvnig}}
\end{figure}
\fi

\subsubsection{Truncated Multivariate Normal Model} \label{sec:tmvn}% page 150 of goodnotes
Next, we place a multivariate normal prior on $\beta$ truncated to the first orthant. In particular, $\beta \sim \m N_d (0, \sigma^2 \lambda^{-1}I_d) \cdot \ind_{[0, \infty)^d}$, where $\sigma^2, \lambda$ are known. This produces a posterior distribution of the form,
\begin{align*}
	& \beta \mid y \sim \m N_d (\beta \mid Q^{-1} b, Q^{-1} ) \cdot \ind_{[0, \infty)^d},
\end{align*}
where $Q, b$ are defined in the Supplement. Then, the marginal likelihood can be written as
\begin{align*}
	p(y) &= \int_R \m N(y \mid X\beta, \sigma^2 I_n) 2^{-d} \m N( \beta \mid 0, \sigma^2 \lambda^{-1}I_d) \ d \beta \\
	&= C \cdot \int_{R} \det(Q)^{\frac{1}{2}}
     e^{ -\frac{1}{2} \left(\beta - Q^{-1} b \right)' Q \left( \beta - Q^{-1}b\right)} d\beta.
\end{align*}
Here, $R = [0,\infty)^\infty$ and $C$ is a known constant term. Note that in this case, however, the integral is not analytically available and prevents the marginal likelihood from being easily computed. \cite{botev_2016} uses a minimax tilting method to calculate the normalizing constant of truncated normal distributions and shows that the proposed estimator has the vanishing relative error property \citep{kroese_taimre_botev_2011}. In light of this, we accept Botev's estimator as the true marginal likelihood in the following experiments. The \texttt{TruncatedNormal} package \citep{tmvn_package} provides samples from truncated normal distributions, so posterior samples from $\beta \mid y$ are readily available. 

In Figure \ref{fig:regression}, we present the simulation results for the case when $d = 20$. Each approximation uses 45 MCMC samples, and we compare the results against the true log marginal likelihood. Once again, the HybE outperforms the other estimators and reinforces its ability to deal with a scarce number of samples. Provided with a sufficiently large number of samples, however, the BSE and CAME are both eventually able produce more accurate results than the HybE.

\begin{figure*}[!ht]
    \centering
    \settoheight{\MyHeight}{\includegraphics[scale=0.49]{example-image-a}}
    \begin{subfigure}[t]{0.49\textwidth}
        \centering
        \includegraphics[scale=0.32]{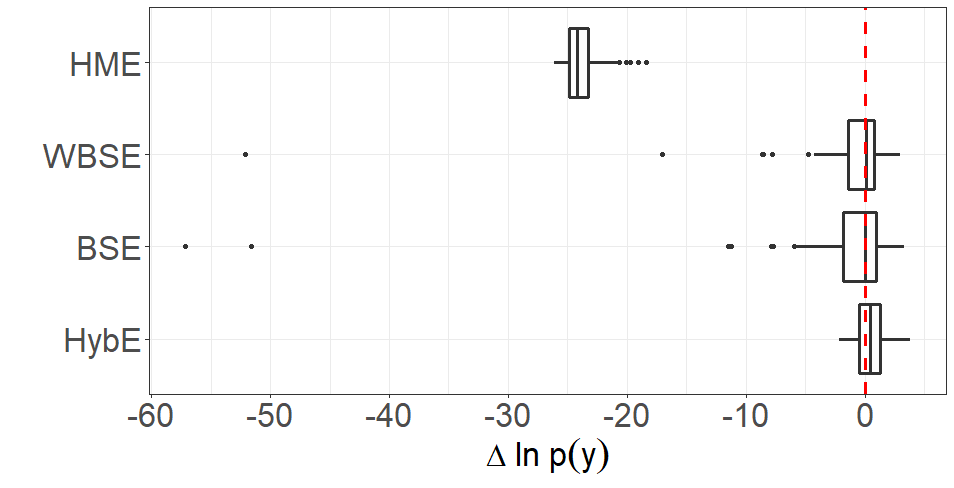}
    \end{subfigure}
    \hfill
    \begin{subfigure}[t]{0.49\textwidth}
        \centering
        \includegraphics[scale=0.32]{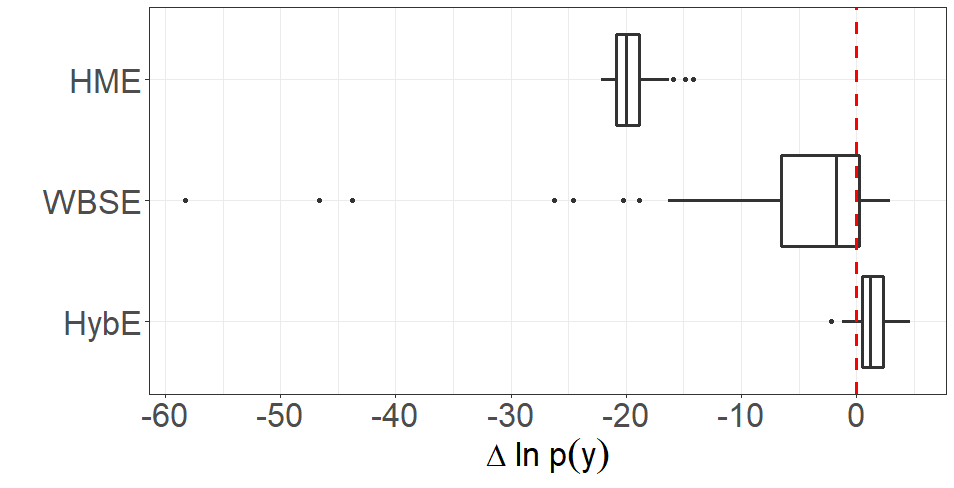}
    \end{subfigure}
    \caption{\textit{Boxplots of the error (truth - estimate) for the unrestricted covariance (left, true $\log p(y)$: -673.7057) and graphical model (right, true $\log p(y)$: -506.3061) examples. Results are reported over 100 simulations, with 100 observations and 25 MCMC samples. For the IW example, we consider $4 \times 4$ covariance matrices with 10 free parameters. For the HIW example, we consider $5 \times 5$ precision matrices with 10 free parameters. Note that we do not include BSE results in the HIW example because the Bridge Sampling algorithm fails to converge with only 25 MCMC samples.} \label{fig:covariance}}
\end{figure*}

\subsection{Unrestricted Covariance Matrices} 
The examples have thus far dealt with parameters in Euclidean space. In the next set of examples, we move beyond the usual Euclidean space and consider parameters in $\mathbb{R}^{d \times d}$. In particular, let $x_1, \ldots, x_n \stackrel{\text{iid}} \sim \m \m N_d(0, \Sigma)$, where $\Sigma \in \mathbb{R}^{d \times d}$. Then the likelihood can be written as follows,
\begin{align} \label{eq:covar_lik}
    & L(\Sigma) = (2 \pi)^{-nd/2} \, \det(\Sigma)^{-n/2} e^{-\tr(\Sigma^{-1} S)/2},
\end{align}
where $S = \sum_{i=1}^n x_i x_i'$. For simplicity, we consider a conjugate inverse-Wishart (IW) prior, $\m W^{-1}(\Lambda, \nu)$, for $\Sigma$, where $\Lambda$ is positive definite $d \times d$ matrix and $\nu >  d - 1$ is the degrees of freedom. Consequently, the posterior distribution of $\Sigma$ is $\m W^{-1}(\Lambda + \mbox{S}, \nu + n)$, and we can compute the marginal likelihood in closed form.

Note that despite being able to sample from the posterior distribution, we cannot yet carry out the Hybrid approximation algorithm. Since posterior samples are drawn from a sub-manifold of $\mathbb{R}^{d\times d}$, if we were to proceed as usual to obtain a partition over $\mathbb{R}^{d\times d}$, there would be no guarantee that a given point within the partition could be used to reconstruct a valid covariance matrix. As such, we circumvent this issue by taking the Cholesky factorization of $\Sigma$, so that $\Sigma = TT'$, where $T$ is a lower triangular matrix with positive diagonal entries, $t_{jj}$ for $j = 1, \ldots, d$. 

Under this transformation, we can define $\Psi(T) = -\log L(T) - \log \pi (T)$, where Eq.\,\eqref{eq:covar_lik} gives us %$L(T) = (2 \pi)^{-nd/2} \, \det(T)^{-n} e^{-\tr( (T T')^{-1} S)/2}$. 
\begin{align*}
    L(T) = (2 \pi)^{-nd/2} \, \det(T)^{-n} e^{-\tr( (T T')^{-1} S)/2}.
\end{align*}
Conveniently, the determinant of the Jacobian matrix $J$ of this transformation is well-known; $|J| = 2^d \prod_{j=1}^d t_{jj}^{d+1-j}$.
%\begin{align*}
 %   |J| = 2^d \prod_{j=1}^d t_{jj}^{d+1-j}.
%\end{align*}
By the change of variable formula, the induced prior on $T$ is 
\begin{align*}
	\pi(T) &= C_{\Lambda, \nu} \, \det(T)^{-(\nu+d+1)} e^{-\tr((T T')^{-1} \Lambda)/2} \  \\
	& \quad \times 2^d \prod_{j=1}^d t_{jj}^{d+1-j},
\end{align*}
where $C_{\Lambda, \nu} = \det(\Lambda)^{\nu/2} / (2^{\nu d/2} \, \Gamma_d(\nu/2))$, and $\Gamma_d(\cdot)$ is the multivariate gamma function. Obtaining posterior samples of $T$ is trivial, as we can simply draw $\Sigma$ from $\m W^{-1}(\Lambda + S, \nu + n)$, and then take the lower Cholesky factor. With this general setup in place, it is worth noting that even with another prior on $\Sigma$, we can carry out the entire algorithm, provided that we have a way to sample from the posterior of $\Sigma$ and a way to compute the Jacobian of the transformation. 

\iffalse
\begin{table}[h]
%\centering
\caption{\textit{Mean and standard deviation of the estimates for the inverse-Wishart model, taken over 100 replications. Here, we consider $4 \times 4$ covariance matrices with 10 free parameters. Each replication has 100 observations and 25 posterior samples. The true log marginal likelihood is -673.706. }}
\begin{center}
\label{tab:coviw_j1e4}
\begin{tabular}{l|c|c|c|c}
\textbf{Method} & \multicolumn{1}{l|}{\textbf{Mean}} & \multicolumn{1}{l|}{\textbf{SD}} & \multicolumn{1}{l|}{\textbf{AE}} & \multicolumn{1}{l}{\textbf{RMSE}} \\ \hline
log HME  & -649.798  & 1.502  &  -23.907 &  23.954 \\
log BSE  & -671.771  & 8.028  &   -1.934 &   8.219 \\
log WBSE & -672.623  & 5.785  &  -1.083  &   5.859 \\
log HybE & -674.153  & 1.142  &    0.447 &   1.221
\end{tabular}
\end{center}
\end{table}
\fi

In Figure \ref{fig:covariance}, we present the results for which each approximation uses 25 MCMC samples. The boxplot of each approximation's errors solidify the robustness of the Hybrid estimator, which produces accurate and low-variance estimates. Although the BSE and WBSE both cover the true log marginal likelihood value, it is apparent that these estimators suffer from stability and convergence issues that are not present in the Hybrid estimator.
%%%%%%%%%% graph algorithms ----------------------------------------------------------------------------------
\subsection{Graphical Models} \label{sec:graphical}
In the following examples, we extend the previous analysis of covariance matrices in the graphical modeling context.  Gaussian graphical models are a popular tool to learn the dependence structure among variables of interest. Consider independent and identically distributed vectors $x_1, x_2, \ldots, x_n$ drawn from a $d$-variate normal distribution with mean vector $0$ and a sparse inverse covariance matrix $\Omega$.  If the variables $i$ and $j$ do not share an edge in a graph $G$, then $\Omega_{ij}=0$. Hence, an undirected (or concentration) graphical model corresponding to $G$ restricts the inverse covariance matrix $\Omega$ to a linear subspace of the cone of positive definite matrices.  A probabilistic framework for learning the dependence structure and the graph $G$ requires specification of a prior distribution for $(\Omega, G)$.  Conditional on $G$, a hyper-inverse Wishart (HIW) distribution \citep{dawid1993hyper} on $\Sigma = \Omega^{-1}$  and the corresponding induced class of distributions on $\Omega$ \citep{roverato2000cholesky} are attractive choices of priors.

\subsubsection{HIW Induced Cholesky Factor Density }\label{sssec:HIWchol}
Denoted by $\mbox{HIW}_G(\delta, B)$, the hyper-inverse Wishart distribution is a distribution on the cone of $d \times d$ positive definite matrices with parameters $\delta > 0$  and a fixed $d \times d$ positive definite matrix $B$. Refer to equations (4) and (5) of \citep{roverato2000cholesky} for the form of the density. When $G$ is decomposable,  an alternative parameterization is given by the Cholesky decomposition $TT'$ of $\Omega = \Sigma^{-1}$ . Provided that the vertices of $G = (V, E)$ are enumerated according to a {\em perfect vertex elimination scheme}, the upper triangular matrix $T'$ has the same zero pattern as $\Omega$. Since the likelihood function is identical to the one given in Eq.\,\eqref{eq:covar_lik}, we need only compute the induced prior on $T$ to complete the definition of $\Psi(T)$. Following \cite{roverato2000cholesky}, the determinant of the Jacobian matrix $J$ of this transformation is given by $|J| = 2^d \prod_{i=1}^d t_{ii}^{\nu_i +1}$, 
where the $i$th row of $T'$ has exactly $\nu_i + 1$ many nonzero elements. More specifically, let $\mbox{ne}(v_i) = \{j: (v_i, v_j) \in E \}$. Then $\nu_i =  |\mbox{ne}(v_i) \cap \{i+1, \ldots, d\}|$. The induced joint density of the elements of $T'$, i.e.,
$t_{sr}$ for $s < r$ with the edge $(v_s, v_r) \in E$, and $t_{ii}, i=1, \ldots, d$, is given by 
\begin{align*}
	\pi(T) &=  \bigg[\prod_{i=1}^d \frac{2^{-(\delta + \nu_i)/2}}{\Gamma((\delta + \nu_i)/2)} \times t_{ii}^{(\delta + \nu_i -2)} e^{-\frac{1}{2} t_{ii}^2} (2t_{ii}) \bigg] \\
	& \quad \times \Bigg[ \prod_{(r,s): r > s, (v_s,v_r) \in E } \frac{1}{\sqrt{2\pi}} e^{-\frac{1}{2} t_{sr}^2}\Bigg].
\end{align*}
Since we are able to sample from the posterior distribution, $\mbox{HIW}_G(\delta + n, B + S)$, where $S = \sum_{i=1}^n x_i x_i'$, we are well-equipped to compute the Hybrid estimator. For this example, we take $\delta = 3$ and $B = I_5$, and in Figure \ref{fig:covariance}, we present the errors for the different estimators when 25 MCMC samples are used for each approximation. Even with limited MCMC samples, the HybE retains its ability to produce reliable results that do not exhibit the high variance that we see in the WBSE. As the number of MCMC samples increases, the WBSE stabilizes and eventually beats the HybE.

\iffalse
\begin{table}[h]
%\centering
\caption{\textit{Mean and standard deviation of the estimates for the Hyper inverse-Wishart model, taken over 100 replications. Here, we consider $5 \times 5$ covariance matrices with 10 free parameters. Each replication has 100 observations and 25 posterior samples. The true log marginal likelihood is -506.306.}}
\begin{center}
\label{tab:hiw_j1e4}
\begin{tabular}{l|c|c|c|c}
\textbf{Method} & \multicolumn{1}{l|}{\textbf{Mean}} & \multicolumn{1}{l|}{\textbf{SD}} & \multicolumn{1}{l|}{\textbf{AE}} & \multicolumn{1}{l}{\textbf{RMSE}} \\ \hline
log HME     & -486.675    & 0.895  & -19.631 & 19.699 \\
log WBSE    & -501.225    & 9.815  & -5.081  & 11.009 \\
log HybE    & -507.760    & 1.362  &  1.454  & 1.988
\end{tabular}
\end{center}
\end{table}
\fi

\iffalse
       approx approx_sd    mae      ae   rmse
LIL  -506.306     0.000  0.000   0.000  0.000
HybE -507.760     1.362  1.603   1.454  1.988
WBSE -501.225     9.815  5.693  -5.081 11.009
HME  -486.675     1.641 19.631 -19.631 19.699
\fi

\subsection{Approximate Posterior Samples} \label{sec:vb}
Up until now, we have assumed that asymptotically exact samples from the posterior distribution are available to be used as input for the proposed approximation. In fact, for all previous numerical experiments, we have used samples drawn from the exact posterior distribution, ridding us of the need for burn-in or thinning. We now investigate how these algorithms perform when we only have approximate posterior samples. As a demonstration, we revisit the MVN-IG example in Section \ref{sec:mvnig-reg} and consider the case where $\beta \in \mathbb{R}^9$. We construct the following mean field approximation to the posterior distribution, $q(\beta, \sigma^2) = q(\beta) q(\sigma^2)$, where
\begin{align*}
q(\beta) \equiv \prod_{i=1}^3 \m N_3 \left( \mu_n^{(i)}, \sigma^2_0 V_{n}^{(i)} \right), \ q(\sigma^2) \equiv \m{IG} \left( a_n, b_n \right).
\end{align*}
Here, we have split the original 9-dimensional normal distribution into a product of 3-dimensional normal distributions, with the mean and covariance components extracted from the true posterior parameters. In particular, $\mu_{n}^{(1)} = (\mu_{n,1}, \mu_{n2}, \mu_{n3})'$, $\mu_{n}^{(2)} = (\mu_{n4}, \mu_{n5}, \mu_{n6})'$, $\mu_{n}^{(3)} = (\mu_{n7}, \mu_{n8}, \mu_{n9})'$. Each $V_n^{(i)}$ is defined as the corresponding $3 \times 3$ block matrix in $V_n$, and $\sigma_0^2$ is the posterior mean of $\sigma^2$.

In Figure \ref{fig:approx_mvnig} below, we observe that even with non-exact posterior samples, the Hybrid approximation produces accurate estimates, with an average error of 0.449 over 100 replications, compared to average errors of 0.698 and 1.035 for the CAME and BSE, respectively. While the latter two estimators have lower variance than the Hybrid approximations, neither covers the true marginal likelihood.  

\begin{figure}[h]
\includegraphics[scale=0.33]{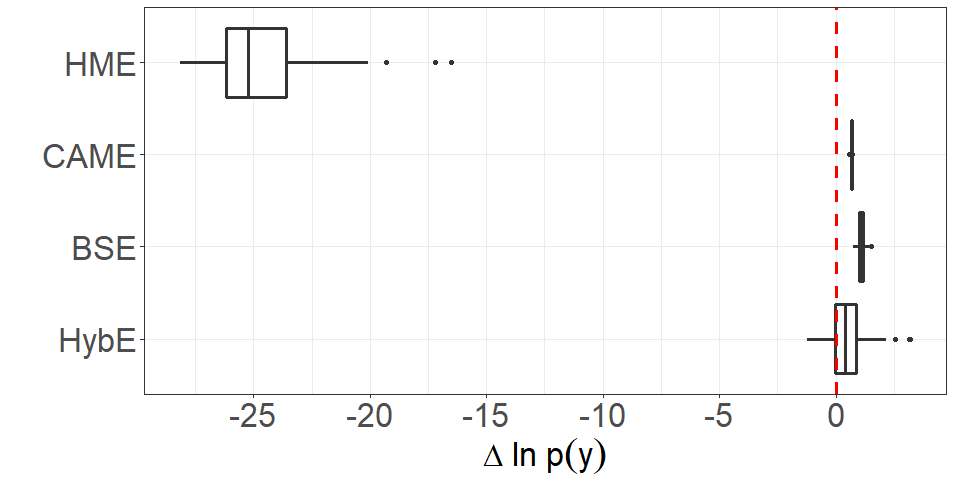}
\caption{\textit{Boxplots of the error (truth - estimate) for the MVN-IG example $(\beta \in \mathbb{R}^9)$ with approximate posterior samples. For each of the 100 replications, we used 100 observations and 100 approximate posterior samples. The true log marginal likelihood is -147.3245.} \label{fig:approx_mvnig}}
\end{figure}

\section{CONCLUSION} \label{sec:conclusion}

In this paper, we developed a novel algorithm that combines a variety of ideas to efficiently estimate the marginal likelihood. By first using a regression tree to identify high probability regions of the parameter space and then leveraging numerical integration ideas to obviate the need to trust the quality of the MCMC samples, we are able to construct an approximation that scales well with both the dimension and the complexity of the parameter space. From the simulation studies, we see that the Hybrid estimator is both accurate and reliable, providing robust approximations in situations when MCMC samples are either scarce or non-exact. Therefore, our contribution is multifaceted and bears practical value such that even in higher dimensions and in instances where generating MCMC samples is expensive and/or non-exact, the Hybrid estimator still delivers promising results.

Furthermore, the Hybrid approximation scheme outlined in this paper lays the groundwork for future work in a number of possible directions. One area of potential refinement is the construction of the partition of the parameter space. While we used CART for its convenience and interpretability, we found that the default objective function for CART was unsuitable for determining the representative point of each partition set, and we had to solve an additional optimization problem to obtain these points. Instead of this two-step roundabout approach, where we used CART to learn the partition and the objective function in Eq.\, \eqref{eq:obj_Q} to identify representative points, we could modify the CART objective function to directly target the desired objective.

Another aspect of the current algorithm that can be further developed is the current formulation of the local approximation to $\Psi$ in each of the partition sets. The piecewise constant approximation in Eq.\,\eqref{eqn:constant_approx}, though providing encouraging results, is a rather simplistic way to approximate $\Psi$, particularly when moving to higher dimensions. A natural extension to the constant approximation is to use a local Taylor expansion to introduce higher order terms, giving piecewise linear and quadratic approximations.

\subsubsection*{Acknowledgements}
The authors are grateful for the anonymous reviewers for providing valuable comments and suggestions. The authors also express special thanks to Donald Chung for insightful discussion about code optimizations.

%%%% references -----------------------------------
\bibliographystyle{plainnat}
\bibliography{hybridml}

\begin{thebibliography}{24}
\providecommand{\natexlab}[1]{#1}
\providecommand{\url}[1]{\texttt{#1}}
\expandafter\ifx\csname urlstyle\endcsname\relax
  \providecommand{\doi}[1]{doi: #1}\else
  \providecommand{\doi}{doi: \begingroup \urlstyle{rm}\Url}\fi

\bibitem[Binev et~al.(2005)Binev, Cohen, Dahmen, DeVore, and
  Temlyakov]{binev_cohen_dahmen_devore_temlyakov_2005}
Peter Binev, Albert Cohen, Wolfgang Dahmen, Ronald DeVore, and Vladimir
  Temlyakov.
\newblock Universal algorithms for learning theory part i: Piecewise constant
  functions.
\newblock \emph{Journal of Machine Learning Research}, 6:\penalty0 1297–1321,
  2005.

\bibitem[Botev(2016)]{botev_2016}
Z.~I. Botev.
\newblock The normal law under linear restrictions: simulation and estimation
  via minimax tilting.
\newblock \emph{Journal of the Royal Statistical Society: Series B (Statistical
  Methodology)}, 79\penalty0 (1):\penalty0 125–148, 2016.
\newblock \doi{10.1111/rssb.12162}.

\bibitem[Botev and Belzile(2019)]{tmvn_package}
Zdravko Botev and Leo Belzile.
\newblock \emph{TruncatedNormal: Truncated Multivariate Normal and Student
  Distributions}, 2019.
\newblock URL \url{https://CRAN.R-project.org/package=TruncatedNormal}.
\newblock R package version 2.1.

\bibitem[Breiman(1984)]{breiman_1984}
Leo Breiman.
\newblock \emph{Classification and regression trees}.
\newblock Wadsworth International Group, 1984.

\bibitem[Chib(1995)]{chib_1995}
Siddhartha Chib.
\newblock Marginal likelihood from the gibbs output.
\newblock \emph{Journal of the American Statistical Association}, 90\penalty0
  (432):\penalty0 1313–1321, 1995.
\newblock \doi{10.1080/01621459.1995.10476635}.

\bibitem[Chib and Jeliazkov(2001)]{chib_jeliazkov_2001}
Siddhartha Chib and Ivan Jeliazkov.
\newblock Marginal likelihood from the metropolis–hastings output.
\newblock \emph{Journal of the American Statistical Association}, 96\penalty0
  (453):\penalty0 270–281, 2001.
\newblock \doi{10.1198/016214501750332848}.

\bibitem[Dawid and Lauritzen(1993)]{dawid1993hyper}
A~Philip Dawid and Steffen~L Lauritzen.
\newblock Hyper markov laws in the statistical analysis of decomposable
  graphical models.
\newblock \emph{The Annals of Statistics}, pages 1272--1317, 1993.

\bibitem[Friel and Wyse(2012)]{friel_wyse_2012}
Nial Friel and Jason Wyse.
\newblock Estimating the evidence - a review.
\newblock \emph{Statistica Neerlandica}, 66\penalty0 (3):\penalty0 288–308,
  2012.
\newblock \doi{10.1111/j.1467-9574.2011.00515.x}.

\bibitem[Ghosal and Van Der~Vaart(2007)]{ghosal2007convergence}
Subhashis Ghosal and Aad Van Der~Vaart.
\newblock Convergence rates of posterior distributions for noniid observations.
\newblock \emph{The Annals of Statistics}, 35\penalty0 (1):\penalty0 192--223,
  2007.

\bibitem[Gronau et~al.(2020)Gronau, Singmann, and Wagenmakers]{bs_package}
Quentin~F. Gronau, Henrik Singmann, and Eric-Jan Wagenmakers.
\newblock {bridgesampling}: An {R} package for estimating normalizing
  constants.
\newblock \emph{Journal of Statistical Software}, 92\penalty0 (10):\penalty0
  1--29, 2020.
\newblock \doi{10.18637/jss.v092.i10}.

\bibitem[Kroese et~al.(2011)Kroese, Taimre, and
  Botev]{kroese_taimre_botev_2011}
Dirk~P. Kroese, Thomas Taimre, and Zdravko~I. Botev.
\newblock \emph{Handbook of Monte Carlo methods}.
\newblock Wiley-Blackwell, 2011.

\bibitem[Lenk(2009)]{lenk_2009}
Peter Lenk.
\newblock Simulation pseudo-bias correction to the harmonic mean estimator of
  integrated likelihoods.
\newblock \emph{Journal of Computational and Graphical Statistics}, 18\penalty0
  (4):\penalty0 941–960, 2009.
\newblock \doi{10.1198/jcgs.2009.08022}.

\bibitem[Meng and Schilling(2002)]{meng_schilling_2002}
Xiao-Li Meng and Stephen Schilling.
\newblock Warp bridge sampling.
\newblock \emph{Journal of Computational and Graphical Statistics}, 11\penalty0
  (3):\penalty0 552–586, 2002.
\newblock \doi{10.1198/106186002457}.

\bibitem[Meng and Wong(1996)]{meng_wong}
Xiao-Li Meng and Wing~Hung Wong.
\newblock Simulating ratios of normalizing constants via a simple identity: a
  theoretical exploration.
\newblock \emph{Statistia Sinica}, 6:\penalty0 831–860, 1996.
\newblock URL \url{www.jstor.org/stable/24306045}.

\bibitem[Neal(2001)]{neal_2001}
Radford~M. Neal.
\newblock Annealed importance sampling.
\newblock \emph{Statistics and Computing}, 11\penalty0 (2):\penalty0 125–139,
  2001.
\newblock \doi{10.1023/a:1008923215028}.

\bibitem[Newton and Raftery(1994)]{newton_raftery_1994}
Michael~A. Newton and Adrian~E. Raftery.
\newblock Approximate bayesian inference with the weighted likelihood
  bootstrap.
\newblock \emph{Journal of the Royal Statistical Society: Series B
  (Methodological)}, 56\penalty0 (1):\penalty0 3–26, 1994.
\newblock \doi{10.1111/j.2517-6161.1994.tb01956.x}.

\bibitem[Pajor(2017)]{pajor_2017}
Anna Pajor.
\newblock Estimating the marginal likelihood using the arithmetic mean
  identity.
\newblock \emph{Bayesian Analysis}, 12\penalty0 (1):\penalty0 261–287, 2017.
\newblock \doi{10.1214/16-ba1001}.

\bibitem[Raftery et~al.(2007)Raftery, Newton, Satagopan, and
  Krivitsky]{raftery_newton_satagopan_krivitsky_2007}
Adrian~E Raftery, Michael~A Newton, Jaya~M Satagopan, and Pavel~N Krivitsky.
\newblock Estimating the integrated likelihood via posterior simulation using
  the harmonic mean identity.
\newblock \emph{Bayesian Statistics}, 8:\penalty0 1–45, 2007.

\bibitem[Rezende and Mohamed(2015)]{pmlr-v37-rezende15}
Danilo Rezende and Shakir Mohamed.
\newblock Variational inference with normalizing flows.
\newblock In Francis Bach and David Blei, editors, \emph{Proceedings of the
  32nd International Conference on Machine Learning}, volume~37 of
  \emph{Proceedings of Machine Learning Research}, pages 1530--1538, Lille,
  France, 07--09 Jul 2015. PMLR.

\bibitem[Roverato(2000)]{roverato2000cholesky}
Alberto Roverato.
\newblock Cholesky decomposition of a hyper inverse wishart matrix.
\newblock \emph{Biometrika}, 87\penalty0 (1):\penalty0 99--112, 2000.

\bibitem[Salimans et~al.(2015)Salimans, Kingma, and
  Welling]{pmlr-v37-salimans15}
Tim Salimans, Diederik Kingma, and Max Welling.
\newblock Markov chain monte carlo and variational inference: Bridging the gap.
\newblock In Francis Bach and David Blei, editors, \emph{Proceedings of the
  32nd International Conference on Machine Learning}, volume~37 of
  \emph{Proceedings of Machine Learning Research}, pages 1218--1226, Lille,
  France, 07--09 Jul 2015. PMLR.

\bibitem[Skilling(2006)]{skilling_2006}
John Skilling.
\newblock Nested sampling for general bayesian computation.
\newblock \emph{Bayesian Analysis}, 1\penalty0 (4):\penalty0 833–859, 2006.
\newblock \doi{10.1214/06-ba127}.

\bibitem[Therneau and Atkinson(2019)]{rpart_package}
Terry Therneau and Beth Atkinson.
\newblock \emph{rpart: Recursive Partitioning and Regression Trees}, 2019.
\newblock URL \url{https://CRAN.R-project.org/package=rpart}.
\newblock R package version 4.1-15.

\bibitem[Tierney and Kadane(1986)]{tierney_kadane_1986}
Luke Tierney and Joseph~B. Kadane.
\newblock Accurate approximations for posterior moments and marginal densities.
\newblock \emph{Journal of the American Statistical Association}, 81\penalty0
  (393):\penalty0 82–86, 1986.
\newblock \doi{10.1080/01621459.1986.10478240}.

\end{thebibliography}

\onecolumn

\aistatstitle{Supplementary Materials}

\section{EXPERIMENTS}

\subsection{Implementation}

As a part of the proposed Hybrid approximation algorithm, we used the CART model implemented in the \texttt{rpart} package \citep{rpart_package} in \texttt{R} for the tree-fitting process. Under the default settings, we obtained the partition of the parameter space from the fitted model. The rest of the Hybrid approximation algorithm, as elicited in Algorithm \ref{alg:hybrid}, is implemented in \texttt{R} and \texttt{C++}. 

\subsubsection{Construction of Competing Methods}
We obtain estimates from a number of competing methods, such as the Harmonic mean estimator (HME), Corrected Arithmetic Mean estimator (CAME), Bridge Sampling estimator (BSE), and Warp Bridge Sampling estimator (WBSE). For the CAME, we use the importance sampling estimator shown in Eq. (19) in \cite{pajor_2017}. Moreover, since the conjugate normal model and multivariate normal inverse-gamma model in Sections \ref{sec:conjugate2d} and \ref{sec:mvnig-reg}, respectively, are very similar to those given in \cite{pajor_2017}, we select the same importance distributions given in the aforementioned paper. For the truncated multivariate normal distribution example in Section \ref{sec:tmvn}, we take the importance distribution to be a truncated multivariate normal distribution with the mean and variance components estimated from the posterior samples. For the BSE and WBSE, we rely on the implementation provided by the \texttt{bridgesampling} package in \texttt{R}; see \cite{bs_package} for more details. 

\subsection{Details of Examples}

Below, we present the calculations and formulae associated with each of the examples that we included in Section \ref{sec:results}. If available in closed form, we include the expressions for the posterior distributions and the marginal likelihoods. In instances where the marginal likelihood cannot be written analytically, we turn to existing packages that provide specialized approximations that have been shown in literature to be accurate. In addition, we provide the true parameter values used to generate the data, as well as the prior hyperparameter settings used to obtain the true log marginal likelihood values.

\subsubsection{Conjugate Normal Model}
In the conjugate normal model in Section \ref{sec:conjugate2d}, the posterior distribution of $(\mu, \sigma^2)$ is well-known. In particular, $\mu \mid \sigma^2, y_{1:n} \sim \m N ( m_n, \sigma^2/w_n)$ and $\sigma^2 \mid y_{1:n} \sim \m{IG} (r_n / 2, s_n / 2)$, with posterior parameters defined as follows, 
\begin{align*}
    m_n &= \frac{m\bar{y} + w_0 m_0}{n + w_0}, \quad w_n = w_0 + n, \quad r_n = r_0 + n, \\
    s_n &= s_0 + \sum_{i=1}^n (y_i - \bar{y})^2 + \left( \frac{nw_0}{n + w_0} \right)(\bar{y} - m_0)^2.
\end{align*}
With this in place, the marginal likelihood can be computed in closed form, 
\begin{align} \label{eq:conjug_ml}
    p(y) = \pi^{-n/2} \left( \frac{w_0}{w_n} \right)^{1/2} \frac{\Gamma\left( \frac{r_n}{2}\right)}{\Gamma \left( \frac{r_0}{2} \right)} \cdot \frac{s_0^{r_0 / 2}}{s_n^{r_n / 2}}.
\end{align}

Each of the $n = 100$ observations was drawn from a normal distribution with mean 30 and variance 4. The prior hyperparameters were $m_0 = 0, w_0 = 0.05, r_0 = 3, s_0 = 3$. Plugging these into Eq. \eqref{eq:conjug_ml}, we computed the true log marginal likelihood to be -113.143. 

\subsubsection{Multivariate Normal Inverse-Gamma} \label{sec:mvnig_supp}
Recall the linear regression setup given in Section \ref{sec:regression} and \ref{sec:mvnig-reg}. Here, the posterior distribution can be shown to be of the following form:
\begin{gather*}
    \beta \mid \sigma^2, y \sim \m N \left( \mu_n, \sigma^2 V_{n} \right), \\
    \sigma^2 \mid y \sim \m{IG} \left( a_n, b_n \right),
\end{gather*}
with posterior parameters $\mu_n = V_n (X'y + V_\beta^{-1} \mu_\beta), V_n = (X'X + V_\beta^{-1})^{-1}$, $a_n = a_0 + n / 2 $, and $b_n = b_0 + (y'y + \mu_{\beta}' V_{\beta}^{-1} \mu_{\beta} - \mu_n' V_n^{-1} \mu_n )$. Then the marginal likelihood can be computed directly to be
\begin{align} \label{eq:logml_mvn_ig}
	& p(y) = \frac{1}{(2\pi)^{n/2}}\frac{b_0^{a_0}}{b_n^{a_n}} \frac{\Gamma(a_n)}{\Gamma(a_0)} \frac{\mathrm{det}(V_n)^{1/2}}{\mathrm{det}(V_\beta)^{1/2}}.
\end{align}

\begin{figure}[!h]
\begin{center}
    \includegraphics[scale=0.7]{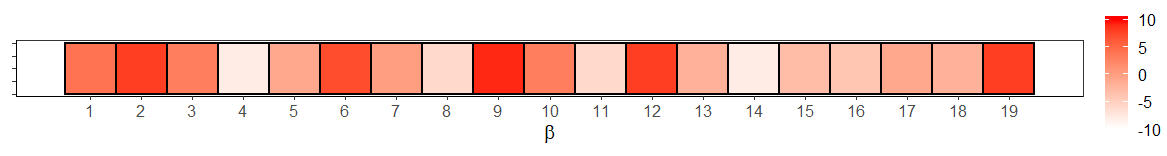}
\caption{\textit{True value of $\beta$; each component is represented as a tile and takes on value between -10 and 10. Values closer to 10 are red and values closer to -10 are white.} \label{fig:mvnig_beta}}
\end{center}
\end{figure}

Each of the 100 observations was drawn from a $d$-dimensional normal distribution according to the linear regression model presented in Section \ref{sec:regression}. In the experiments, we took $d = 19$, and the prior hyperparameters were $\mu_{\beta} = 0_d, V_{\beta} = I_d, a_0 = 1, b_0 = 1$. The true value of $\beta$ is shown as a heatmap in Figure \ref{fig:mvnig_beta} and $\sigma^2 = 4$. Plugging these into Eq. \eqref{eq:logml_mvn_ig}, we computed the true log marginal likelihood to be -303.8482. 

\subsubsection{Truncated Multivariate Normal}

\begin{figure}[!h]
\begin{center}
    \includegraphics[scale=0.7]{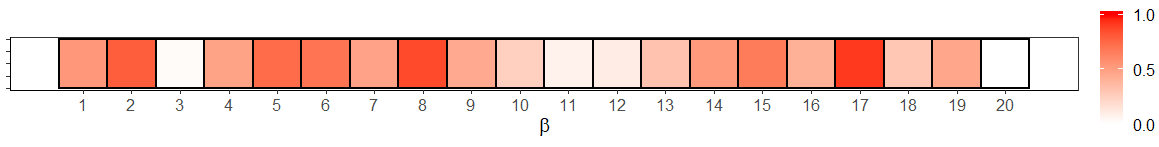}
\caption{\textit{True value of $\beta$; each component is represented as a tile and takes on value between 0 and 1. Values closer to 1 are red and values closer to 0 are white.} \label{fig:tmvn_beta}}
\end{center}
\end{figure}

With the truncated multivariate normal prior given in Section \ref{sec:tmvn}, we obtain the following form of the posterior distribution of $\beta$,
\begin{align*}
    \beta \mid y \sim \m N_d (\beta \mid Q^{-1} b, Q^{-1} ) \cdot \ind_{[0, \infty)^d},
\end{align*}
with posterior parameters $Q = \frac{1}{\sigma^2}(X'X + \lambda I_d)$ and $b = \frac{1}{\sigma^2} X'y$. Each of the $n = 100$ observations was drawn from a $d$-dimensional normal distribution according to the linear regression model presented in Section \ref{sec:regression}. In the experiments, we took $d = 20$, and the prior hyperparameters were $\sigma^2 = 4, \lambda = 0.25$. The true value of $\beta$ is shown as a heat map in Figure \ref{fig:tmvn_beta}. Due to the intractable marginal likelihood, we used the \texttt{TruncatedNormal} package to compute the true marginal likelihood to be -250.2755. 

\subsubsection{Unrestricted Covariance Matrices}

We consider the inverse-Wishart prior on $\Sigma$, $\m W^{-1}(\Lambda, \nu)$, where $\Lambda$ is a positive definite $d \times d$ matrix, and $\nu > d - 1$. The prior density has the following form,

\begin{align*} % \label{eq:IW}
& \pi(\Sigma) = C_{\Lambda, \nu} \, \det(\Sigma)^{-(\nu+d+1)/2} e^{-\tr(\Sigma^{-1} \Lambda)/2},  
\end{align*}
where $C_{\Lambda, \nu} = \det(\Lambda)^{\nu/2} / (2^{\nu d/2} \, \Gamma_d(\nu/2))$. Here, $\Gamma_d(\cdot)$ is the multivariate gamma function, given by
\begin{align*}
	& \Gamma_d(a) = \pi^{d(d-1)/4} \prod_{j=1}^d \Gamma \left( a + (1-j)/2 \right),
\end{align*}
where $\Gamma(\cdot)$ is the ordinary gamma function. Our choice of the prior admits the following closed form marginal likelihood:
\be \label{eq:iw_logml}
\int L(\Sigma) \pi(\Sigma) d \Sigma = \frac{ \Gamma_d((n+\nu)/2) }{\pi^{nd/2} \Gamma_d(\nu/2)} \, \frac{ \det(\Lambda)^{\nu/2}}{ \det(\Lambda + \mbox{S})^{(n+\nu)/2}}. 
\ee

In each of the 100 replications, we took $d = 4$ and drew $n = 100$ observations from a 4-dimensional normal distribution with mean vector $0_d$ and covariance matrix $\Sigma$, where
\begin{equation*}
\Sigma = 
    \begin{bmatrix}
    \phantom{-}1.662  & \phantom{-}1.640  & -1.985            & -0.007\\
    \phantom{-}1.640  & \phantom{-}7.163  & -4.146            &  \phantom{-}5.654\\
    -1.985            & -4.146            &  \phantom{-}4.906 & -1.237\\
    -0.007            & \phantom{-}5.654  & -1.237            &  \phantom{-}6.779 
    \end{bmatrix}    
\end{equation*}

The prior hyperparameters were $\Lambda = I_4$ and $\nu = 5$. Plugging these into Eq. \eqref{eq:iw_logml}, we computed the true log marginal likelihood to be -673.7057.

\subsubsection{Hyper-Inverse Wishart Induced Cholesky Factor Density}

We first introduce some notation to help us obtain a closed form for the marginal likelihood of a decomposable graph $G$. For an $n \times d$ matrix $X$, $X_C$ is defined as the submatrix of $X$ consisting of columns with indices in the clique $C$. Let $(\mathrm{x}_1, \mathrm{x}_2, \ldots, \mathrm{x}_d)=(x_1,x_2,\ldots,x_n)'$, where $\mathrm{x}_i$ is the $i$th column of $X_{n\times d}$. If $C = \{ i_1, i_2, \ldots, i_{\abs{C}} \}$, where $1\leq i_1 < i_2 < \ldots < i_{\abs{C}} \leq d$, then $X_C = (\mathrm{x}_{i_1}, \mathrm{x}_{i_2}, \ldots, \mathrm{x}_{i_{\abs{C}}})$. For any square matrix $A = {(a_{ij})}_{d \times d}$, define $A_C = {(a_{ij})}_{\abs{C}\times\abs{C}}$ where $i, j \in C$, and the order of entries carries into the new submatrix $A_C$. Therefore, $X_C'X_C = (X'X)_C$. 

Decomposable graphs correspond to a special kind of sparsity pattern in $\Sigma$, henceforth denoted $\Sigma_G$. Suppose we have a $\mathrm{HIW}_G (b, D)$ distribution on the cone of $d\times d$ positive definite matrices with $b > 2$ degrees of freedom and a fixed $d \times d$ positive definite matrix $D$ such that the joint density factorizes on the junction tree of the given decomposable graph $G$ as
\begin{equation}\label{eq:HIW1}
	p(\Sigma_G \mid b, D) = \frac{\prod_{C \in \mathcal{C}} p(\Sigma_C \mid b, D_C)}
	{\prod_{S \in \mathcal{S}} p(\Sigma_S \mid b, D_S)},
\end{equation}
where for each $C \in \mathcal{C}$, $\Sigma_C \sim \mbox{IW}_{\abs{C}}(b, D_C)$ with density 
\begin{equation}\label{eq:HIW2}
	p(\Sigma_C \mid b, D_C) \propto \abs{\Sigma_C}^{-(b +2 \abs{C})/2} \mbox{etr}\Big\{-\frac{1}{2}\Sigma_C^{-1} D_C\Big\},
\end{equation}
where $\abs{C}$ is the cardinality of the clique $C$ and $\mbox{etr}(\cdot)=\exp\big\{\mbox{tr}(\cdot)\big\}$. $\mbox{IW}_d(b,D)$ is the inverse-Wishart distribution with degrees of freedom $b$ and a fixed $d\times d$ positive definite matrix $D$ with normalizing constant 
\begin{equation*}
	\abs{\frac{1}{2}D}^{(b+d-1)/{2}}\Gamma^{-1}_d\Big(\frac{b+d-1}{2}\Big).
\end{equation*} 
Note that we can establish equivalence to the parametrization used in Section \ref{sssec:HIWchol} by taking $\delta = b + d - 1$. Since the joint density in Eq. \eqref{eq:covar_lik} factorizes over cliques and separators in the same way as in \eqref{eq:HIW1}-\eqref{eq:HIW2},
\begin{equation}
	f(X\mid\Sigma_G) = {(2\pi)}^{-\frac{np}{2}} \frac
	{\prod_{C\in\mathcal{C}} {\abs{\Sigma_C}}^{-\frac{n}{2}} \mbox{etr}\Big( -\frac{1}{2}\Sigma_C^{-1}\mathrm{X}_C'\mathrm{X}_C \Big)}
	{\prod_{S\in\mathcal{S}} {\abs{\Sigma_S}}^{-\frac{n}{2}} \mbox{etr}\Big( -\frac{1}{2}\Sigma_S^{-1}\mathrm{X}_S'\mathrm{X}_S \Big)}. \label{eq:likelihood}
\end{equation}

The $\mbox{HIW}(b,D)$ density can be written as 
\begin{align*}
	f(\Sigma_G\mid G) &= \frac{\prod_{C\in\mathcal{C}} p(\Sigma_C\mid b, D_C)}{\prod_{S\in\mathcal{S}} p(\Sigma_S\mid b, D_S)} \\
	&= \frac{\prod_{C\in\mathcal{C}} \abs{\frac{1}{2}D_C}^{\frac{b+\abs{C}-1}{2}}\Gamma^{-1}_{\abs{C}}\big(\frac{b+\abs{C}-1}{2}\big)\abs{\Sigma_C}^{-\frac{b+2\abs{C}}{2}} \mbox{etr}\big( -\frac{1}{2} \Sigma_C^{-1}D_C \big)}
	{\prod_{S\in\mathcal{S}} \abs{\frac{1}{2}D_S}^{\frac{b+\abs{S}-1}{2}}\Gamma^{-1}_{\abs{S}}\big(\frac{b+\abs{S}-1}{2}\big)\abs{\Sigma_S}^{-\frac{b+2\abs{S}}{2}} \mbox{etr}\big( -\frac{1}{2} \Sigma_S^{-1}D_S \big )}.
\end{align*}

Then, it is straightforward to obtain the marginal likelihood of the decomposable graph $G$,
\begin{equation}\label{eq:ML}
	f(\mathrm{X}\mid G) = 
	{(2\pi)}^{-\frac{np}{2}} \frac{h(G,b,D)}{h(G,b+n,D+S)} 
	= {(2\pi)}^{-\frac{np}{2}} \frac{\prod_{C\in\mathcal{C}} w(C) }{\prod_{S\in\mathcal{S}} w(S)},
\end{equation}

where
\begin{equation*}
	h(G,b,D) = \frac{\prod_{C\in\mathcal{C}}\abs{\frac{1}{2}D_C}^{\frac{b+\abs{C}-1}{2}}\Gamma^{-1}_{\abs{C}}\big(\frac{b+\abs{C}-1}{2}\big)} {\prod_{S\in\mathcal{S}}\abs{\frac{1}{2}D_S}^{\frac{b+\abs{S}-1}{2}}\Gamma^{-1}_{\abs{S}}\big(\frac{b+\abs{S}-1}{2}\big)}, \, \quad 
	w(C) = \frac{\abs{D_C}^{\frac{b+\abs{C}-1}{2}} {\abs{D_C+\mathrm{X}_C'\mathrm{X}_C}}^{-\frac{b+n+\abs{C}-1}{2}}} {2^{-\frac{n\abs{C}}{2}}\Gamma_{\abs{C}}\big(\frac{b+\abs{C}-1}{2}\big)\Gamma_{\abs{C}}^{-1}\big(\frac{b+n+\abs{C}-1}{2}\big)}.
\end{equation*}

\begin{figure}[!h]
\begin{center}
    \includegraphics[scale=0.5]{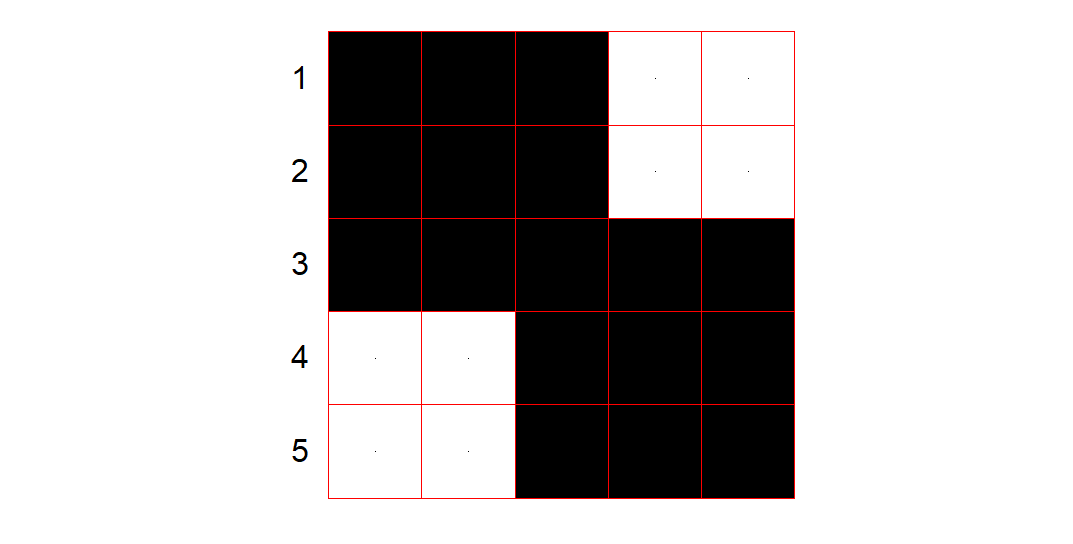}
\caption{\textit{In the undirected graph $G$, with vertex set $V = \{1,2,3,4,5\}$, the $(i,j)$-th box is black if the corresponding edge is present in $G$ and white otherwise.} \label{fig:hiw_graph}}
\end{center}
\end{figure}

Conditional on the graph $G$, represented in Figure \ref{fig:hiw_graph}, we considered a hyper-inverse Wishart prior on $\Sigma = \Omega^{-1}$, $\mathrm{HIW}_G(\delta, B)$, where the prior hyperparameters were $B = I_5$ and $\delta = 3$. We then drew $n = 100$ observations from a $5$-dimensional normal distribution with mean vector 0 and a sparse inverse covariance matrix $\Omega$, where the dependence structure in $\Omega$ was specified by the graph $G$.

Using the formula for the marginal likelihood derived in Eq. \eqref{eq:ML}, we computed the true log marginal likelihood to be -506.3061.

\subsubsection{Approximate Posterior Samples}

For the multivariate normal inverse-gamma example in Section \ref{sec:vb} where we drew approximate posterior samples from the mean field approximation of the posterior distribution, each of the $n = 100$ observations was first drawn from a $d$-dimensional normal distribution according to the linear regression model presented in Section \ref{sec:regression}. In the experiments, we took $d = 9$, and the prior hyperparameters were $\mu_{\beta} = 0_d, V_{\beta} = I_d, a_0 = 1, b_0 = 1$. The posterior distribution of $\beta$ was approximated by a product of 3-dimensional normal distributions, each with mean and covariance components, $(\mu_{n}^{(i)}, V_{n}^{(i)})$ for $i = 1,2,3$. These were extracted from the true posterior parameters $(\mu_n, V_n)$ (defined in Section \ref{sec:mvnig_supp}) in the following way:

\[
\mu_n = \left[ 
\begin{array}{c}
 \mu_{n1} \\ \mu_{n2} \\ \vdots \\ \mu_{n9}
\end{array} \right] = 
\renewcommand{\arraystretch}{1.5}
\left[ 
\begin{array}{@{}c@{}}
 \mu_{n}^{(1)} \\ \hline  \mu_{n}^{(2)} \\ \hline  \mu_{n}^{(3)}
\end{array}
\right], \quad 
V_n =
\left[
\begin{array}{c|c|c}
V_{n}^{(1)} & &  \\
\hline
 & V_{n}^{(2)} &  \\
\hline
 &  & V_{n}^{(3)}
\end{array}
\right]
\]

\begin{figure}[!h]
\begin{center}
    \includegraphics[scale=0.7]{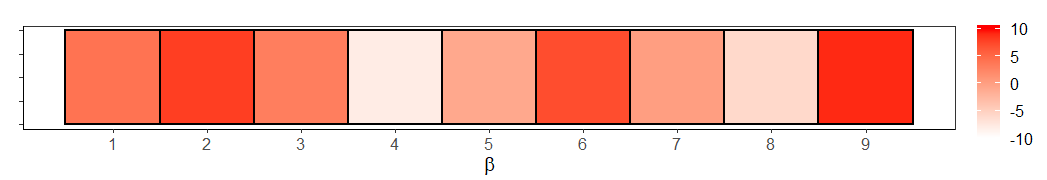}
\caption{\textit{True value of $\beta$; each component is represented as a tile and takes on value between -10 and 10. Values closer to 10 are red and values closer to -10 are white.} \label{fig:vb_beta}}
\end{center}
\end{figure}

The true value of $\beta$ is shown as a heat map in Figure \ref{fig:vb_beta} and $\sigma^2 = 4$. Using the formula for the marginal likelihood derived in Eq. \eqref{eq:logml_mvn_ig}, we computed the true log marginal likelihood to be -147.3245.

\end{document}